# (Meta) Competences for Digital Practice: Educational Scenarios for the Workplace of the Future Exemplified by Building Information Modeling Work Processes

Sebastian Damek
Fachhochschule Erfurt
sebastian.damek@fh-erfurt.de

Heinrich Söbke
Bauhaus-Universität Weimar
heinrich.soebke@uni-weimar.de

Franziska Weise
Chamber of Architects of Thuringia
weise@architekten-thueringen.de

Maria Reichelt
Fachhochschule Erfurt
maria.reichelt@fh-erfurt.de

## ABSTRACT

Workplaces of the future require advanced competence profiles from employees, not least due to new options for teleworking and new complex digital tools. The acquisition of advanced competence profiles is to be addressed by formal education. For example, the method of Building Information Modeling (BIM) aims at digitizing the design, construction and operation of structures and as such requires advanced competence profiles. In this study, two educational scenarios based on teleworking and complex digital tools are compared, each with one cohort and consisting of two learning activities. The first cohort initially completes as first learning activity a semester-long course that aims at BIM domain competences. The semester-long course of the second cohort fosters meta competences, such as communication, collaboration, and digital literacy. At the end of the semester, both cohorts solve in a second learning activity a BIM practice task. Research questions are: (1) Do the two educational scenarios promote the competences to be addressed? And related: (2) What is the impact of the initial course that fosters domain competences or meta competences? Methodologically, the learning outcomes are assessed by measuring the domain competences three times during the educational scenario using online tests in the two cohorts (N=11). Further, students' perceptions are surveyed in parallel using online questionnaires. In addition, semi-structured interviews are conducted at the end of the educational scenarios. The quantitative and qualitative results of the study–designating the training of meta competencies partly as a substitute for imparting domain competences–are presented. Further, the influence of both educational scenarios on competence development for the workplace of the future characterized by telework and digital tools is discussed.

**Keywords**
skill, meta skill, communication, collaboration, digital literacy

## 1 INTRODUCTION

Workplaces and their tasks assigned are changing constantly (Rohrbach-Schmidt and Tiemann 2013), for example due to advancing digitization (Cijan et al. 2019). For meeting the demands of workplaces, employees require competences: workplace competences largely define the capabilities to perform particular tasks. Competences thus have a high significance, for example, for economic prosperity or for social imbalance (Alabdulkareem et al. 2018).

### 1.1 Competences

Literature holds numerous definitions for the concept of competence. Rychen & Salganik (2000) define "the concept of competence refers to the ability to meet demands of a high degree of complexity". Weinert (2001) distinguishes in an OECD report several definitions of competence but comes to the basic–and in this study utilized–conclusion that competence is "considered as a learned, cognitive demand-specific performance disposition [...]". In contrast, Le Deist & Winterton (2005) define competence as a multidimensional construct including the dimensions cognitive competence, functional competence, social competence and meta competence. Furthermore, there is extensive interdisciplinary discourse on definitions and types of competences, including (Rychen and Salganik 2001, 2003; Hartig and Klieme 2006).

### 1.2 Competence Profiles

In professional life, complex competences, which comprise single competences, are frequently required. These complex competences are described systematically by competence models or competence profiles. Competence profiles are used in various ways: for example, competence profiles are a basis for training and study curricula (Mahasneh and Thabet 2015) as well as for designing technology-supported educational scenarios (Niegemann et al. 2008). Changing requirements to competence profiles necessitates changes to formal educational scenarios correspondingly. Furthermore, competence profiles are applied for competence diagnostics (Shavelson 2010), e.g., for the precise matching of workplaces: this involves a comparison between the competence profile demanded by the workplace and the competences of the applicants.

The specification of competence profiles for particular workplaces has been the subject of professional discourse for several decades (e.g., Davis and Miller (1996) or Nugraha et al. (2020)). Amongst others, competence profiles have been specified for hospital leaders (Chung-Herrera et al. 2003), for particular competence domains of workplaces, such as intercultural competences of counselors (Glockshuber 2005), or serious game facilitation competences of instructors (Baalsrud Hauge et al. 2021), for specific competence domains in general, such as entrepreneurship (Arafeh 2015), or for country-specific competences of workplace roles, such as project managers in the construction industry in Poland (Dziekoński 2017). Specifications of competence profiles are often provided by professional organizations, such as the International Project Management Association (IPMA) competence profile for project managers (Lukianov et al. 2019).

In summary, every workplace requires a competence profile. Most competence profiles have changed significantly in recent years due to digitization, e.g., teleworking, and advanced digital tools. Changed competence profiles have to be addressed by modified formal educational scenarios.

### 1.3 Building Information Modeling

One field of work that has evolved greatly in recent years through digitization is Building Information Modeling, a method for digitizing the design, construction, and operation of structures (Sacks et al. 2018). BIM requires the emergence of new competence profiles. For example, the competence profiles of *BIM manager* and *BIM coordinator* are described (Egger et al. 2014). BIM manager and BIM coordinator require technically oriented management competences, such as the implementation of BIM strategies as well as the coordination of BIM processes. To impart BIM competence profiles, country-specific training plans are described, for example, for the Domenican Republic (Rodriguez et al. 2017). In New Zealand, development of BIM competences in the bachelor's degree program Construction is addressed (Rehman et al. 2018). Furthermore, educational programs from professional associations exist, such as buildingSMART, a leading non-profit organization promoting BIM (bSI 2021), or the Australian Institute of Architects (Succar et al. 2012). An overview of educational programs is provided by Rodriguez et al. (2017).

### 1.4 Study Aim

Higher education institutions, as places of formal education, are in charge of imparting competences that students prepare for workplace. In the following, two formal educational scenarios imparting BIM competences are described. The study aim is investigating the effectiveness of the educational scenarios and, more specifically, studying the relevance of meta competences to BIM competences.

## 2    METHODOLOGY

The study was conducted as a pilot field study. Two elective courses, each consisting of one theory and one practice lecture per week, were conducted at two higher education institutions. Both the practice and the theory lectures were held completely online. Firstly, this was a course BIM (n=6), open to master students in *Architecture*, on BIM basics. This course introduced students to BIM-based modeling. The overarching goal of the course, to which all educational activities were aligned, was to generate a 3D building model while respecting the design specifications. For this purpose, students used a complex digital environment using the learning management system moodle (Moodle.org 2018), which integrated the software essential for task completion (including IFCWebServer (Concerted Solutions 2021)) as well as a repository for the data. The building model was to be exportable in accordance with the specifications of the BIM standard data format *Industry Foundation Classes (IFC)* and the data exchange as described by buildingSMART Benelux (2021). The classification scheme to be used was based on Richter and Liedtke (2021). The second course, VirtuIng (n=5), promoted meta competences using a multiplayer online game (Authors 2021) to bachelor students in *Civil Engineering*. This course also was aligned at an overarching goal: students were to learn how to operate the multiplayer online game (CCP 2012) and then work in groups to each find a strategic game objective, such as specialization within the game. Along the process of achieving the game objective, students had to work collaboratively to overcome challenges, which demanded students exercising their meta competences. At the completion of both courses in the winter semester of 2022, the students faced a joint BIM practice task: in three lectures, the students had to check a given building model using a previously unknown BIM validation software and to record issues found. The competences promoted by the courses and the final joint BIM practice task are summarized in Table 1.

*Table 1 Learning goals (competences) of the two courses*

| Course BIM | Course VirtuIng |
|---|---|
| *Domain competences* <ul><li>*Basic knowledge of a CAD software in the domain of 3D modeling.*</li><li>*Basic spatial understanding*</li><li>*Knowledge of IFC as a data exchange format*</li><li>*Modeling of a 3D building model based on planning specifications*</li><li>*Ensuring IFC compatibility of the model*</li></ul> | *Meta competences, such as* <ul><li>*Communication*</li><li>*Collaboration*</li><li>*Critical Thinking*</li><li>*Creativity*</li><li>*Information Literacy*</li></ul> |
| **BIM practice task "Building Model Check"** ||
| <ul><li>*Understanding of complex BIM software*</li><li>*Understanding of file management (including files for models, classifications and checking rules)*</li><li>*Understanding of IFC data structures and of IFC checking rules*</li><li>*Performing a model check using BIM software*</li><li>*Performing a clash check using BIM software*</li><li>*Clear, reflected presentation of the check results.*</li></ul> ||

After an introductory presentation on the BIM practice task and the software (Solibri Inc. (2021)), the instructors were available during the lectures for clarifying any inquiries. The completion of the practice task was not subject to any organizational constraints. However, results were to be individually submitted and presented to two instructors at a time in a 10-minute presentation that was graded.

The study aimed to answer the following research questions: RQ 1: Do the two educational scenarios promote the workplace competences to be addressed? and related: RQ 2: What is the impact of the initial educational scenario that fosters either domain competences or meta competences?

Multiple data collections took place during this study: First, BIM competences were assessed using a learning management system-based test, in which 5 randomly selected single-choice questions were asked from a pool of 15 questions in total. Second, students answered a questionnaire with a total of 19 items on a 5-point Likert scale on workplace competences. Both the test and the questionnaire were collected at three points in time for ascertaining a possible progression over time: At the beginning of the course (Measurement 1), at the beginning of the practice task (Measurement 2), and at the end of the practice task (Measurement 3). Finally, a semi-structured interview was conducted online via video chat with each student after the final presentation. The interviews lasted 10 to 15 minutes, were recorded and later transcribed and analyzed qualitatively by two authors (Schmidt 2004). In case of discrepancies in the evaluations, a common understanding was established in a joint debriefing of both authors.

## 3    RESULTS

### 3.1    Demography

Of the total 11 participants in the study, who were on average just under 22 years old, 9 were male and 2 female. 2 were studying *Computer Science* in their bachelor's degree, 3 *Civil Engineering* in their bachelor's degree and 6 *Architecture* in their master's degree.

### 3.2    Learning

The BIM tests show a similar level of performance in both cohorts (Table 2). The roughly equal results despite more domain-specific learning content in the BIM cohort might be explained by VirtuIng participants being more likely to be IT-savvy than the BIM participants from the discipline of *Architecture*. Overall, there seems to be an increase in knowledge respectively the same level of knowledge is reproduced in less time. A seemingly paradoxical behavior was observed in the BIM cohort: the greater the competences in using the software for 3D modeling was, the lower was the willingness to engage with the BIM practice task. Here, the difficulties of the practice task were apparently underestimated. The results of the final presentation also show only small differences in the assessment (Table 3). Overall, the level of learning outcomes appears nearly similar.

*Table 2. BIM Test*

| Cohort | BIM (n=6) | | VirtuIng (n=5) | |
|---|---|---|---|---|
| **Measurement** | Correct [%] | Time used [s] | Correct [%]t | Time used [s] |
| **Measurement 1** | 35 | N/A | 36 | 159 |
| **Measurement 2** | 40 | N/A | 48 | 136 |
| **Measurement 3** | 55 | N/A | 48 | 120 |

*Table 3. Final Presentation*

| Cohort | BIM (n=6) | VirtuIng (n=5) |
|---|---|---|
| **Results** | 91 % | 87 % |

## 3.3    Questionnaire

The questionnaire comprised three groups: *Task Attractiveness* (5 items), *Learning* (7 items), and *Self-Organization* (7 items). All items are recorded on a 5-point Likert scale (with 1: disagreement and 5: full agreement). The following sections present the results of the three groups, differentiated by cohort and by the three measurements.

### 3.3.1 Task Attractiveness

*Task Attractiveness* was recorded to provide a basis for evaluating further outcomes: the students' valuation of the task to be solved also influences their endeavors to complete the task (Wigfield and Eccles 2000). The results (Figure 1) indicate satisfactory scores for challenge (item A), a higher-than-average interest (B), the perceived need to pay full attention to the tasks (C) and the assessment of own knowledge as rather too low (E). In comparison, the BIM cohort shows a rather higher technical interest (B), a lower challenge (A), and a higher ability to self-motivate. The VirtuIng cohort lacks domain knowledge (E).

*Figure 1. Items regarding task attractiveness and cohort (n=11, mean from 3 measurements)*

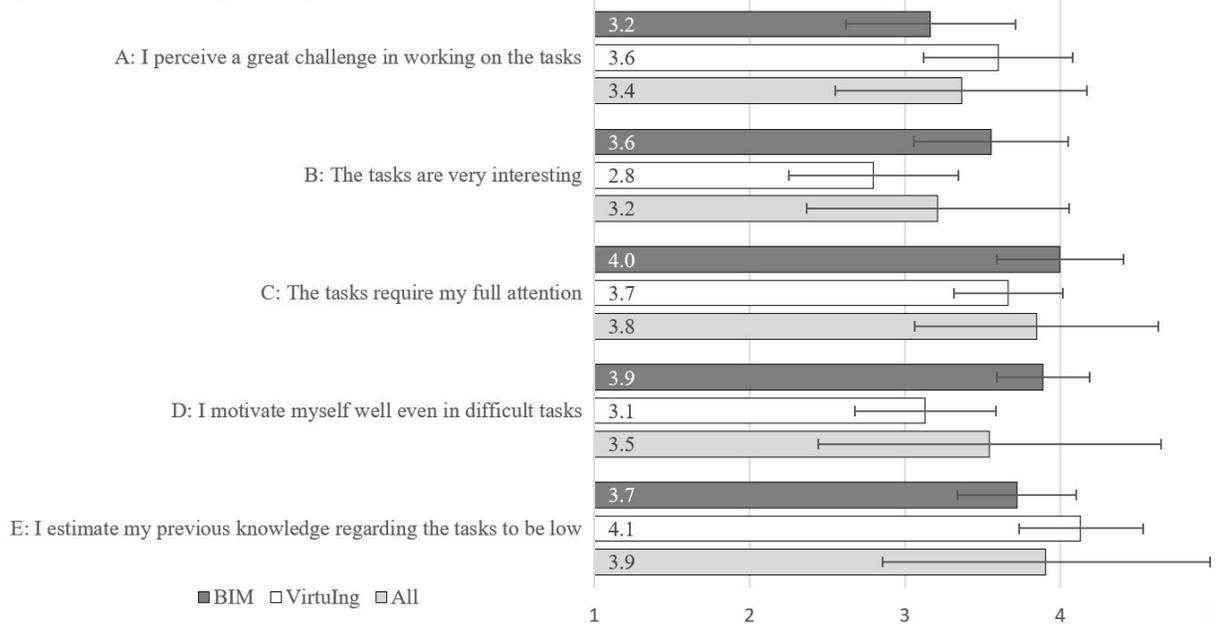

Considering the scores over time (Figure 2), the perceived challenge (A) and the required attention (C) increase with time, while the interest (B) decreases. These opposing developments might be reflected in the decreasing capacity for self-motivation (D). The decrease of missing knowledge (E) from Measurement 2 to Measurement 3 might indicate learning.

*Figure 2. Items regarding task attractiveness and measurements (n=11)*

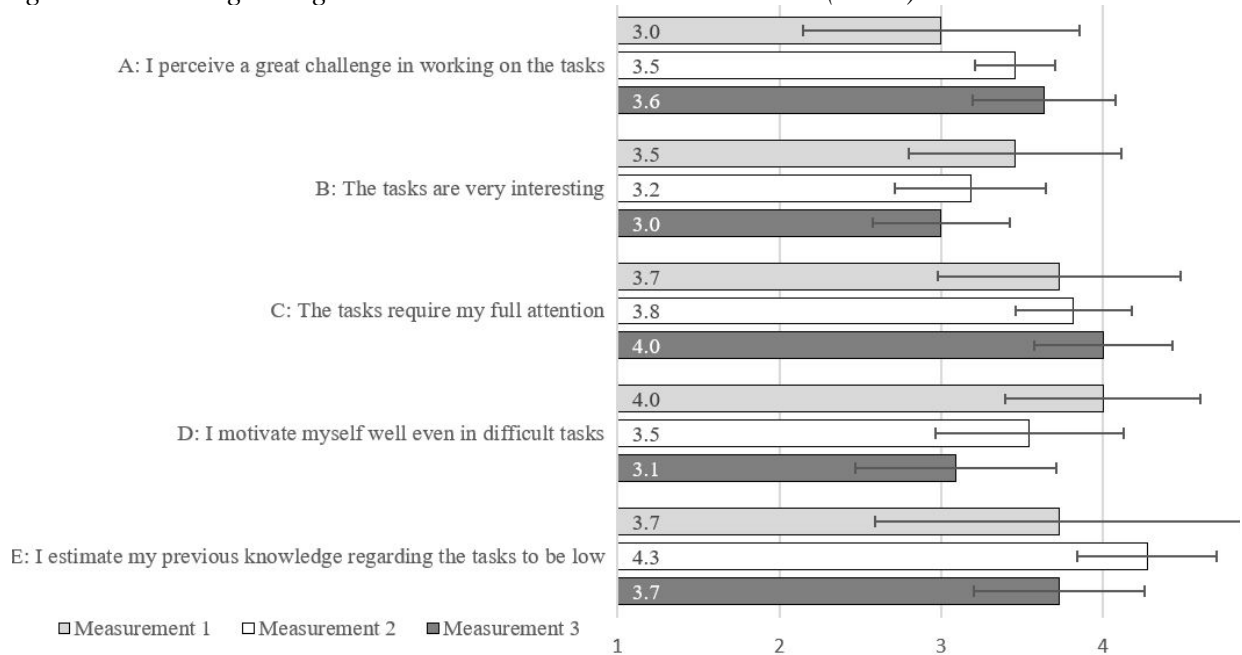

### 3.3.2 Learning

Seven items were combined in the group *Learning*, which captured the students' assessment of the characteristics of the educational scenarios under consideration; larger values connoted positive learning outcomes in each case. Notably, all items were rated above the mean. Comparing the two cohorts (Figure 3), the BIM cohort received higher scores for transfer of knowledge (F and G), which might be due to the more concrete domain learning tasks in the BIM cohort. The same rationale might be applicable to the BIM cohort's higher scores of the learning environment (I). This appears slightly reversed for media literacy (H) and information literacy (J and K), each of which are dedicated learning goals of the VirtuIng cohort. Item L captures perceived self-efficacy; in this respect, the VirtuIng cohort achieved higher scores than the BIM cohort.

*Figure 3. Items regarding learning and cohort (n=11, mean from 3 measurements)*

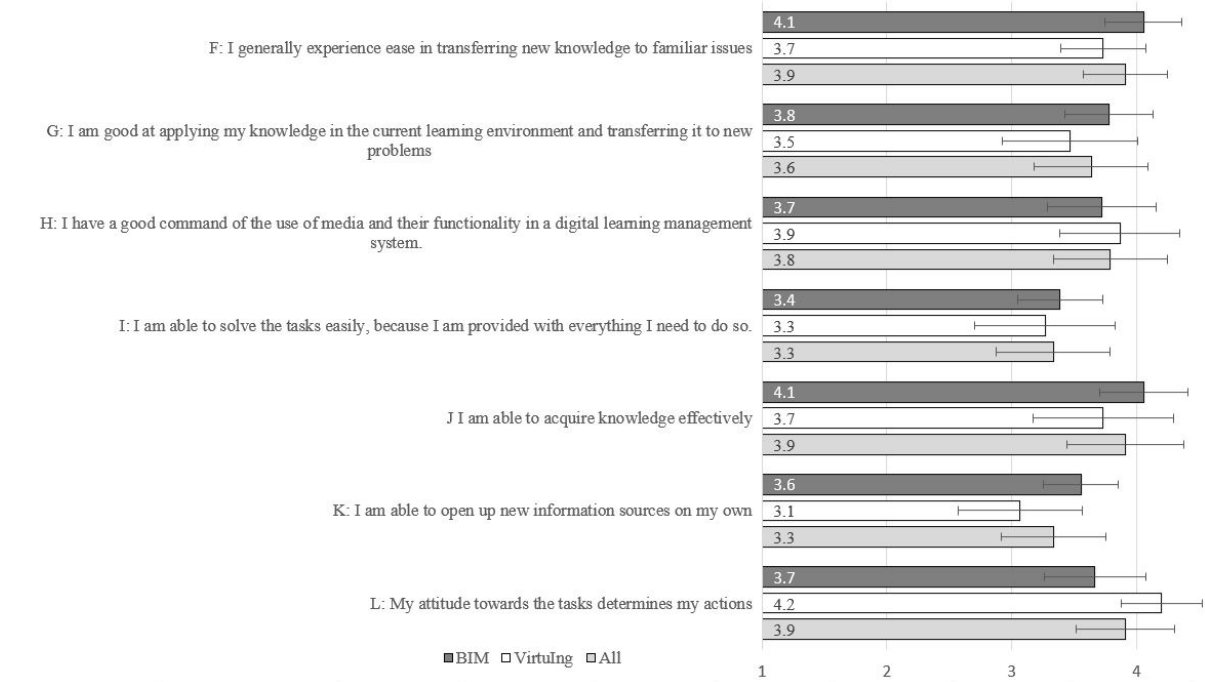

Figure 4 describes the temporal development of the items. For knowledge transfer competences (F and G), the highest scores are found at the beginning. For the other two measurements, the scores drop to almost identical levels. An explanation suggested is that the estimation at the start of the semester is based more on abstract expectations, whereas the last two measurements might be under the impression of the actual domain contents provided in the course. A similar effect might be responsible for the continuous drop of almost one point in the assessment of the learning environment (I), which was changed for measurements 2 and 3. The impression of the new software could be the reason for the drop in media literacy (H) at Measurement 2; Measurement 3 finally shows the highest value of all three measurements. The requirements of the new software and the experience of using it might also not have been beneficial to self-efficacy (L), which drops slightly across all three measurements. A similar effect may be seen for information literacy (J and K).

*Figure 4. Items regarding learning and measurements (n=11)*

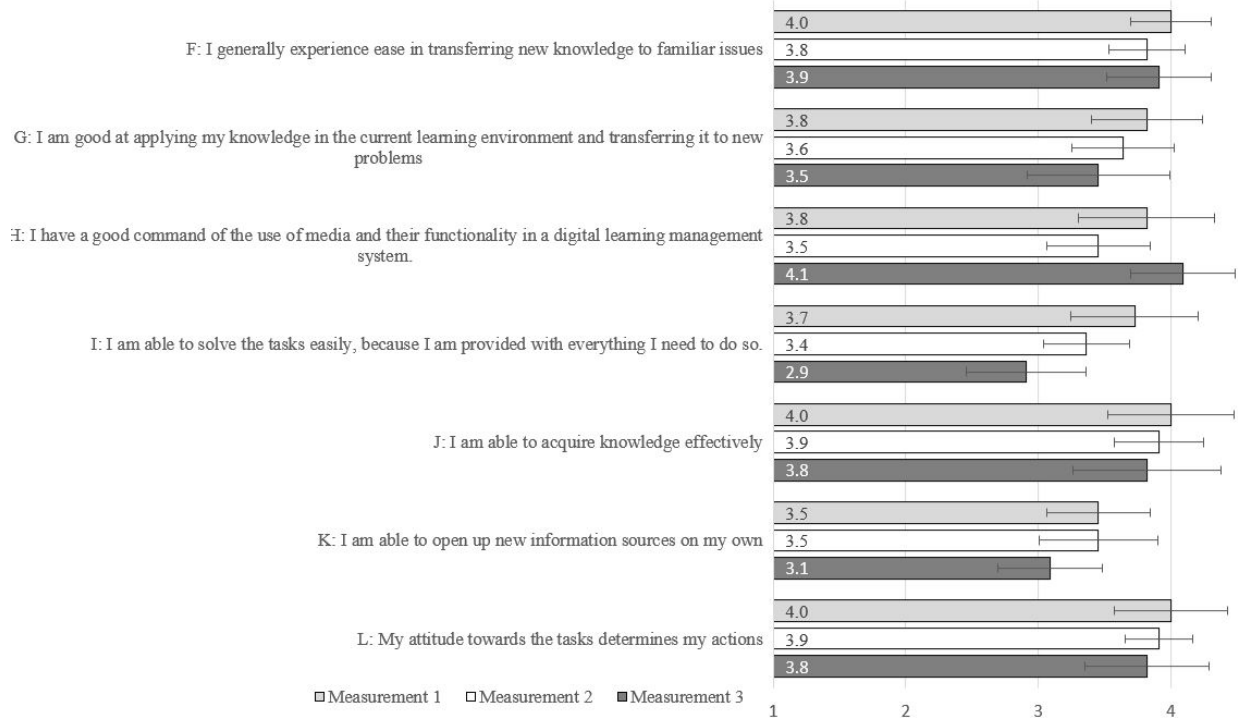

### 3.3.3 Self-Organization

The group *Self-Organization* contained seven items aiming at understanding the abilities of self-organization with the goals of learning and task completion. Also in this group, higher scores indicate positive results. When comparing the two cohorts (Figure 5), there was only one item that received scores in the middle or below: prior knowledge (Q), which tended to be rated as not sufficient, with higher scores in the BIM cohort. It is striking that the BIM cohort also received significantly higher scores for self-organization in learning situations (M) and appears more satisfied with the solution of their tasks (N). An explanation might be the more graspable tasks of BIM modeling than the more abstract tasks of developing meta competences. For the meta competence of communication (O and P), the VirtuIng cohort scores higher than the BIM cohort in each case. The same is true for the competence or willingness to collaborate (R and S). Both meta competences are among the learning goals of the VirtuIng cohort.

*Figure 5. Items regarding self-organization and cohort (n=11, mean from 3 measurements)*

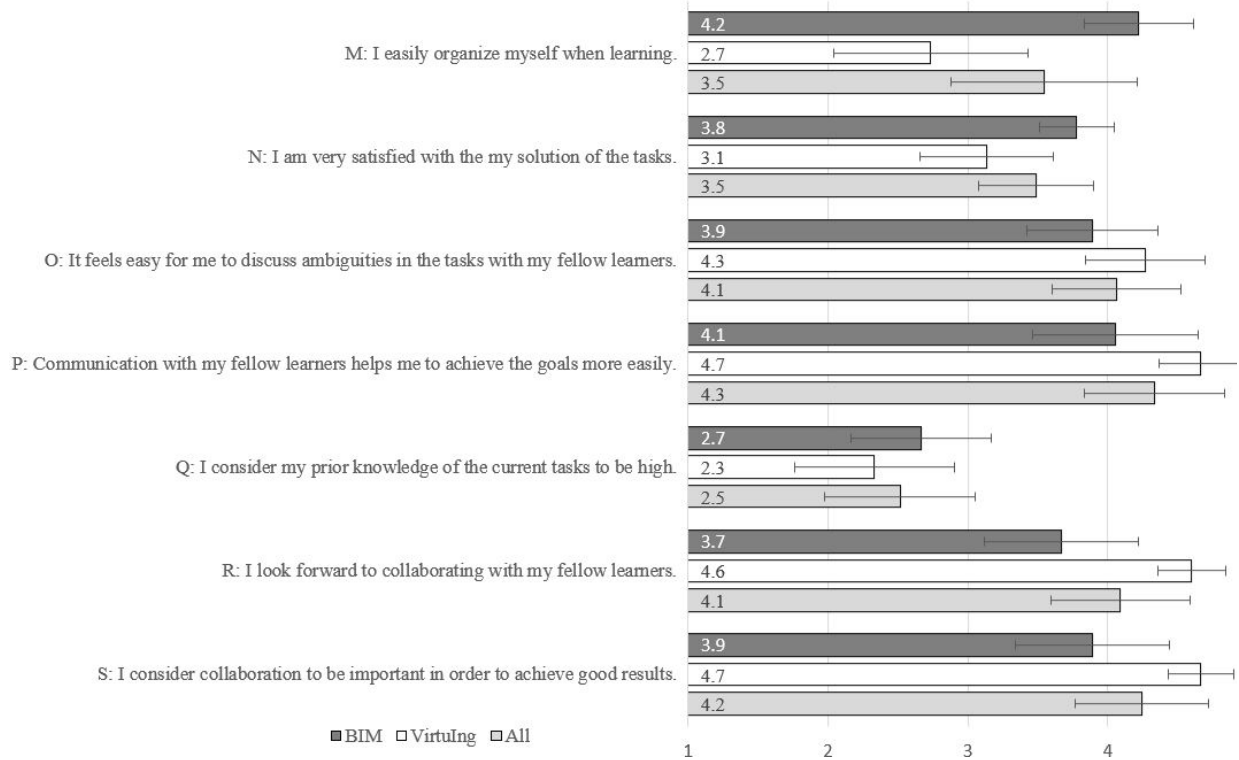

The temporal progression of the scores for self-organization (*Figure 6*) shows decreasing scores for self-organization (M) and also a decreasing satisfaction (N) with the solutions reached. This development may again indicate a rather abstractly grounded perception at the beginning of the semester, which is then increasingly exposed to the challenges of reality during the semester. Thus, perceived prior knowledge (Q) is estimated to be quasi-constant over the semester with a small drop in Measurement 2 at the beginning of the validation task. The opportunity to communicate (O) with fellow students increases slightly over the semester, while the influence of communication to solve the task (P) is seen as more constant. Also perceived as quasi-stable at a comparatively high level is the solution supportiveness of collaboration (R and S).

*Figure 6. Items regarding self-organization and measurements (n=11)*

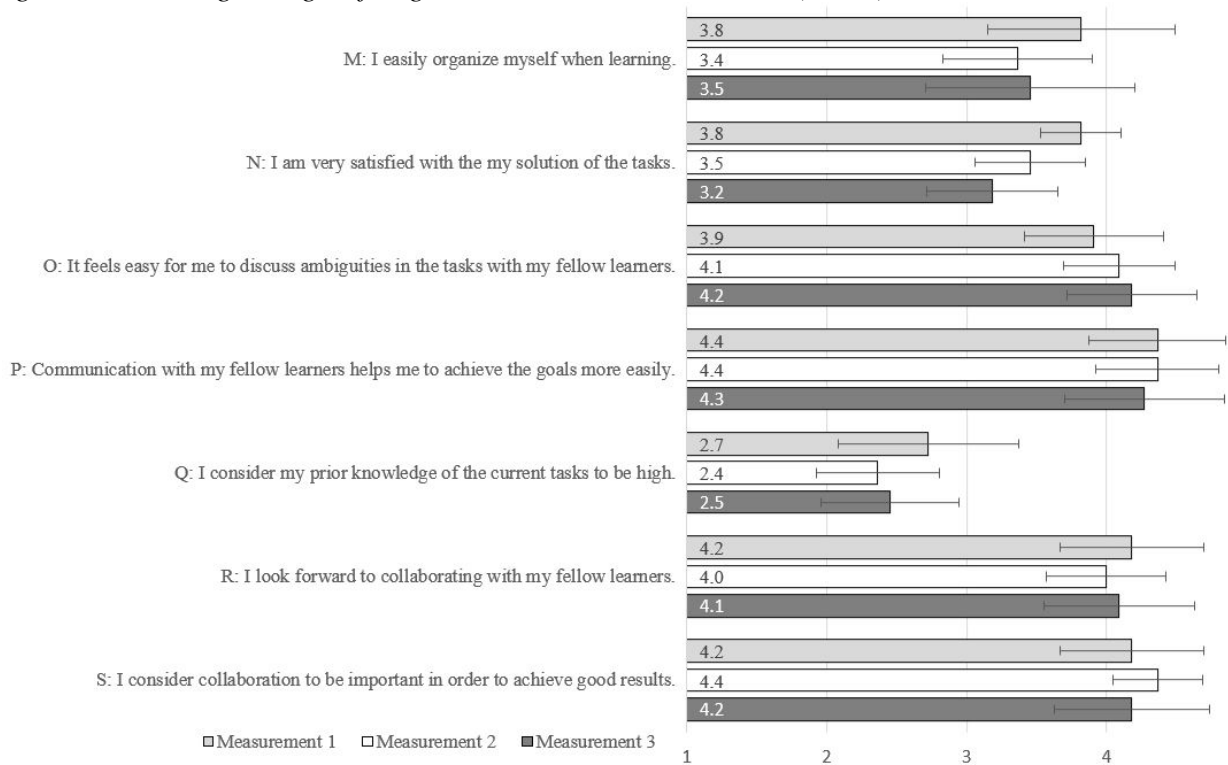

### 3.4 Semi-Structured Interviews

The interviews were structured into a total of six themes, which are summarized below. On average, approximately 4 themes were addressed in each interview. A total of 123 statements were categorized, per theme these were 8 to 37 statements.

### 3.4.1 Task Evaluation

In the first theme of 37 statements categorized, the basic attitude of the students to the course and the tasks to be solved was queried. Most frequently (15 times), it was confirmed that the course brought concrete added value for the students, which was specifically named: the added values included both domain competences such as basic BIM knowledge, understanding of complex software, geometric modeling, and partially automated model checking based on rules, as well as meta competences such as enhanced collaboration. Positive feedback that the task was pleasing and matched the students' preferences was given a total of 10 times. Negative feedback was provided four times, such as frustrating moments due to excessive demands and high time requirements. However, such feedback cannot be considered entirely negative, since partial overstrain is part of learning processes (Vygotsky 1978). Five times there were concrete suggestions for improvement, which referred to the buildings to be modeled, the software to be used, and the didactic support. Two students stated that they had no thematic relations to the task, once existing prior knowledge was mentioned.

### 3.4.2 Satisfaction and Challenges

The second theme containing 17 statements was intended to focus on the students' satisfaction with completed tasks as well as the perceived challenges. Almost all students (10 out of 11) stated being satisfied with the solutions delivered. Challenges included the rather difficult software handling (3 statements) and 2 times each the feeling of not having acquired profound knowledge finally and not having mastered the modeling process.

### 3.4.3 Effects of the Course

The objective of the third theme, with a total of 16 statements, was the students' assessment of the effects of the course taken (i.e., either BIM or VirtuIng) on their ability to solve the final practice task. In particular, differences in assessments by cohort affiliations were expected here. Overall, seven students indicated the first part of the course having provided little to no preparation. Four students, on the other hand, saw parallels through familiarization with complex software, which was also necessary in the first part of the course. Interestingly, all of these statements originated from the VirtuIng cohort that had to familiarize themselves with complex game software (CCP 2012) in the first part. While two other statements found the introduction to the practice task to be positive, two further statements requested improvements (Better integration and video recording of the introductory lecture).

### 3.4.4 Practical Relevance

It was also an objective of the interview to ascertain the extent to which the tasks to be worked on prepare students for later professional practice. A total of eight statements were made that basically acknowledged the imparting of competences relevant to practice, albeit with limitations. Among the limitations were the preparation not being sufficient for a larger context, for the management of a corresponding project or for an in-depth knowledge of the software used.

### 3.4.5 Competences Demanded

For addressing the research questions, the students' assessment of competences particularly required for solving the tasks is crucial. In total, students named specific competences 28 times. Most frequently, this was the ability to collaborate (Ten times), followed by personal competences, such as perseverance, systematic approach to work, accepting imperfect work, the ability to learn and not comparing oneself with others to maintain one's own motivation. Information literacy, i.e., the ability to gather and evaluate relevant information, was also mentioned six times. A total of five times, meta competences were mentioned generally.

### 3.4.6 Digital Competences

Digital competences were specifically covered in the interview due to the research questions. In a total of 17 statements, the enormous importance of competences for handling digital tools was emphasized. In three statements, domain competences, i.e., knowledge of BIM, were considered to be very valuable. In two statements, on the other hand, domain competences were valued less strongly, but rather knowledge of the concrete software was emphasized as relevant.

## 4    DISCUSSION

In both courses, BIM as well as VirtuIng, taking place in predominantly digitized environments, learning outcomes could be identified. Therefore, both courses represent educational scenarios for the transfer of competences, which are largely relevant in the context of the digitization of workplaces (RQ 1). A comparison of the two courses as preparation for a BIM practice task corroborates the importance of meta competences. While the BIM course teaches domain competences of Building Information Modeling, VirtuIng trains meta competences. With the BIM practice task, the VirtuIng cohort had to solve a task without knowing the domain-specific BIM basics, which had to be acquired during the problem solving process. It was evident that the students of the VirtuIng cohort were able to especially advantage from the enhanced competences of collaboration and communication during the practice task. Due to the lack of domain-specific competences, the task completion level of the BIM cohort was not fully achieved. However, remarkably, the students of the VirtuIng cohort did not let themselves be deterred by a task comparatively unfamiliar to them, but approached the task with confidence and ultimately achieved a satisfactory outcome, as indicated by the presentation assessments. In addition, the competences taught in VirtuIng are likely to be particularly useful for management-oriented BIM competence profiles, such as BIM coordinator.

This pilot field study inherently carries some limitations. These include the low number of participants, which prevents results from being representative. Furthermore, the smaller number of participants, especially in the VirtuIng course, may have fostered a well-cohesive group unlike courses with larger numbers of participants. In addition, the implementation of a BIM practice task in VirtuIng is not necessarily practice-oriented, since there is no contextual connection regarding the other course activities, but has the character of a laboratory study. Also, the first measurement covered a time when the VirtuIng cohort probably knew about a practice task to be solved but had not received any domain input on BIM. The corresponding measurements, especially the assessment of BIM knowledge, are therefore subject to a certain degree of inaccuracy. Further, the study only explored a small subset of workplace competences required in BIM processes. Hence, imparting workplace competences in future needs to be systemized and complemented.

## 5    CONCLUSION

The increasing digitization of workplaces leads to altered competence profiles that need to be addressed by formal education. Accordingly, this pilot study described two educational scenarios on domain competences and meta competences and investigated their effect on students' ability of solving a BIM practice task. On the one hand, the results reveal that the objectives of the respective learning scenarios were achieved to an acceptable extent. On the other hand, the results suggest that missing domain competences might be (at least partially) substituted by enhanced meta competences. This finding emphasizes the importance of imparting meta competences in formal education. Overall, this study highlights the option and necessity of adapting formal educational scenarios to empower employees for workplaces being subject to advancing technological evolutions.